\documentclass[aps,reprint,superscriptaddress,showpacs,showkeys,amsmath]{revtex4-1}

\usepackage{graphicx}
\usepackage[colorlinks=true,citecolor=blue,linkcolor=blue,urlcolor=black]{hyperref}

\begin{document}

% Use the \preprint command to place your local institutional report
% number in the upper righthand corner of the title page in preprint mode.
% Multiple \preprint commands are allowed.
% Use the 'preprintnumbers' class option to override journal defaults
% to display numbers if necessary
%\preprint{}

\title{Probing Chiral Electronic Excitations in Bilayer Graphene by Raman scattering}

\author{Elisa \surname{Riccardi}}
\email[Corresponding author: ]{elisa.riccardi@lpa.ens.fr}
\altaffiliation[Present address: ]{Laboratoire Pierre Aigrain, ENS,  24 rue Lhomond, 75005 Paris}
\affiliation{Laboratoire Mat\'eriaux et Ph\'enom\`enes Quantiques (UMR 7162 CNRS), Universit\'e Paris Diderot-Paris 7, B\^atiment Condorcet, 75205 Paris Cedex 13, FR}

\author{Oleksiy \surname{Kashuba}}
\email[Corresponding author: ]{okashuba@physik.uni-wuerzburg.de}
\affiliation{Department of Theoretical Physics and Astrophysics, University of  W\"urzburg, Am Hubland, D-97074 W\"urzburg, Germany}

\author{Maximilien \surname{Cazayous}}
\author{Marie-Aude \surname{M\'easson}}
\altaffiliation[Present address: ]{Institut NEEL CNRS/UGA UPR2940,25 rue des Martyrs BP 166, 38042 Grenoble cedex 9}
\author{Alain \surname{Sacuto}}
\author{Yann \surname{Gallais}}
\email[Corresponding author: ]{yann.gallais@paris7.jussieu.fr}
\affiliation{Laboratoire Mat\'eriaux et Ph\'enom\`enes Quantiques (UMR 7162 CNRS), Universit\'e Paris Diderot-Paris 7, B\^atiment Condorcet, 75205 Paris Cedex 13, FR}

\date{\today}

\begin{abstract}
We report a symmetry resolved electronic Raman scattering (ERS) study of a back-gated bilayer graphene device.
We show that the ERS continuum is dominated by interband chiral excitations of $A_{2}$ symmetry and displays a characteristic Pauli-blocking behavior similar to the monolayer case.
Crucially, we show that non-chiral excitations make a vanishing contribution to the Raman cross-section due to destructive interference effects in the Raman amplitude matrix elements.
This is in a marked contrast to the optical absorption measurements and opens interesting prospects for use of Raman scattering as a selective probe for the detection of the chiral degrees of freedom in graphene, topological materials, and other 2D crystals.
\end{abstract}



\maketitle

\section{Introduction}

Monolayer graphene exhibits an unique low energy electronic band structure which mimics two-dimensional massless Dirac spectrum.
%This unusual dispersion has important consequences for the properties if electrons and holes: monolayer graphene electrons can be described as massless relativistic particles.
Electrons in graphene, similar to the relativistic Dirac particles, have a chiral nature, which has profound consequences for the transport properties of Dirac materials, such as Klein tunneling~\cite{Katsnelson2006}, half-integer quantum Hall effect~\cite{Jiang2007}, weak antilocalization~\cite{Mccann2006b}, etc.
In monolayer graphene, the chirality phenomenon is linked to the existence of two inequivalent sub-lattices, which act as an isospin degree of freedom~\cite{Semenoff1984,Haldane1988}.
Historically, chirality in graphene refers to the projection of the isospin on the direction of momentum, a property, which in the particle physics is called helicity (see Appendix 1 for a discussion of the relation between chirality and helicity in the context of graphene).
Interestingly, chirality can also be defined for bilayer graphene, where low-energy electronic excitations mimic massive Dirac particles. Here, in contrast to monolayer graphene, the chiral nature of the excitations is not related to the sub-lattice degree of freedom, but rather to the index of atomic layer, which can also act as an isospin degree of freedom~\cite{Mccann2006,Mccann2006a,Mccann2013}.
The chirality phenomenon in graphene systems is shared by topological insulators, where strong spin-orbit coupling leads to spin-momentum locking similar to one in Dirac hamiltonian~\cite{Hasan2010}.
The chiral spin-textures at the surface of the topological insulator Bi$_2$Se$_3$ have been successfully probed by Spin and Angle Resolved Photoemission Electron Spectroscopy measurements~\cite{Xia2009,Hsieh2009,Xu2011}, and their associated collective chiral excitations have been recently revealed by Raman spectroscopy~\cite{Kung2017}.

Since chirality in graphene is associated with the sub-lattice index instead of the real spin, accessing the chirality and isospin of the excitations in graphene-like systems has been proved somewhat more elusive.
Electronic Raman scattering (ERS) has been recently emerged as a tool to study electronic excitations in graphene.
The ERS studies were focused on probing of inter-Landau level excitations, which require strong magnetic fields~\cite{Berciaud2014,Faugeras2011,Faugeras2015}. Recently polarization resolved measurements showed that the ERS spectrum at zero magnetic field is dominated by the interband chiral excitations across the Dirac point~\cite{Riccardi2016}.
Here we use a commonly accepted terminology, in which the term ``chiral Raman excitation'' denotes the Raman-excited electron-hole pair, where electron and hole have the opposite chiralities. As predicted theoretically~\cite{KashubaFalko2009}, these excitations display Pauli-blocking behavior upon tuning the Fermi level with a gate voltage.
However, due to its simple band structure, the only available vertical interband transitions in monolayer graphene are chiral, and the ERS spectrum of these interband excitations does not differ significantly from the well-studied infrared absorption spectra~\cite{Basov2014}.
By contrast, in bilayer graphene the absorption experiments reveal a much richer spectrum where only subset of all available excitations are chiral, providing an appealing platform to demonstrate the selectivity of polarization resolved ERS measurements with respect to conventional infrared transmission measurements.

\par
In this paper we demonstrate this selectivity by studying ERS in a bilayer graphene device.
First we provide a theory showing that the dominant ERS processes correspond to the excitations belonging the $A_2$ representation of the lattice symmetry point group  of bilayer graphene.
All other processes are strongly suppressed due to destructive interference effects in the Raman amplitude matrix elements.
The dominant ERS processes require an isospin flip, i.e. they result in the creation of chiral Raman-active electron-hole excitations, and represent only a subset of all interband transitions in bilayer graphene: namely, the interband transitions that are mirror-symmetric with respect to the charge neutral point. These predictions are tested experimentally by investigating the symmetry resolved ERS spectrum of a bilayer graphene device under varying gate voltage.
The ERS continuum is dominated by interband excitations of $A_{2}$ symmetry and display a characteristic Pauli-blocking behavior upon varying gate voltage, which can be reproduced by taking into account chiral excitations only.
This demonstrates that non-chiral excitations make a vanishing contribution to the Raman cross-section, as predicted theoretically. Concluding, we contrast this unique property of ERS to infrared absorption measurements, which probe all interband transitions, partially disguising the contribution arising from the chiral excitations. 

\section{Raman scattering electronic excitations in bilayer graphene: theory}

\begin{figure}
\includegraphics[width=\columnwidth]{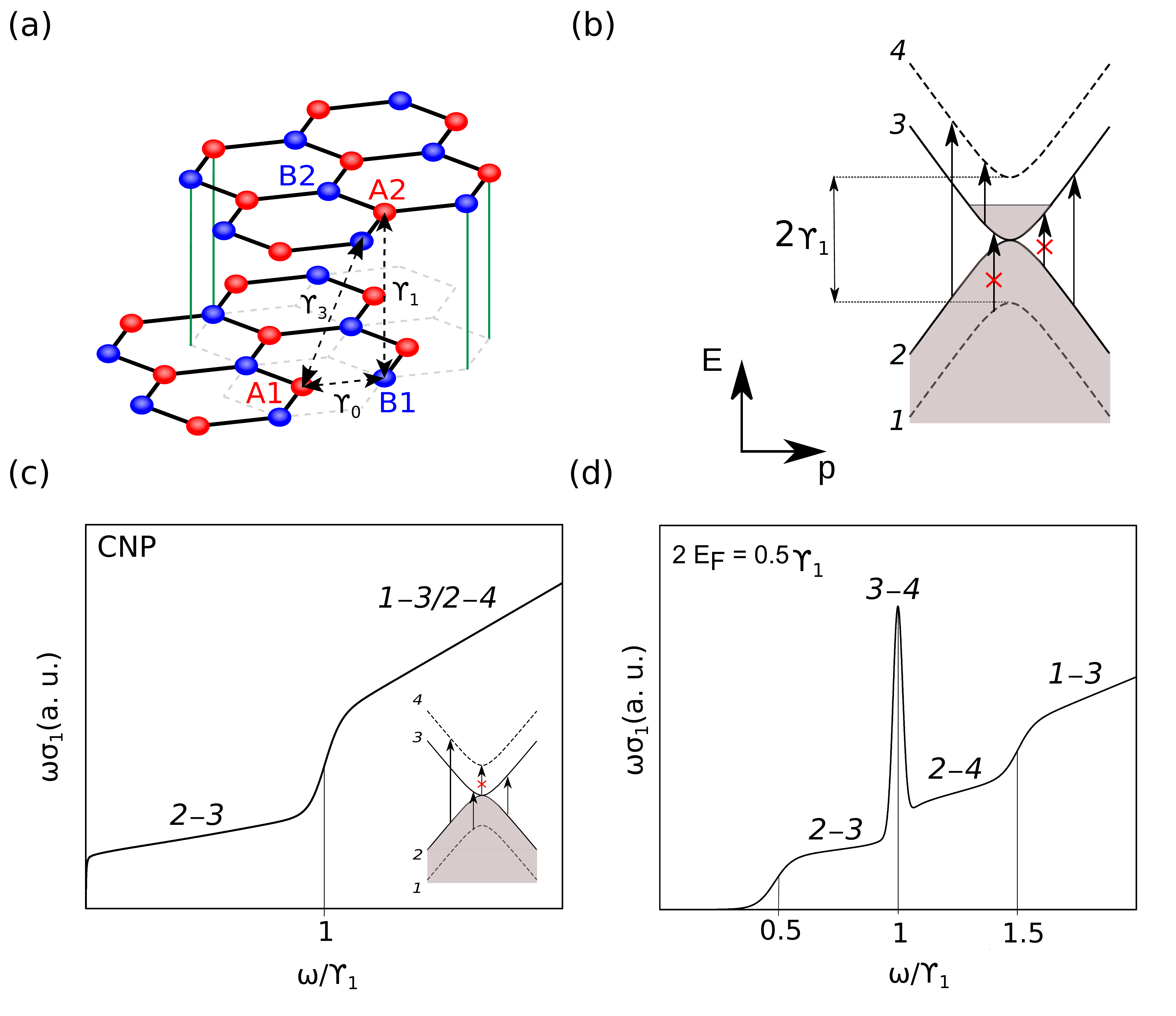}
\caption{ \label{Fig1}(a) Bernal-stacked bilayer graphene.
The unit cell contains four non-equivalent sites, $A1$, $B1$, $A2$ and $B2$.
The atoms $A2$ and $B1$ (dimer atoms) are sitting directly on top of each other.
The hopping amplitudes $\gamma_0$, $\gamma_1$ and $\gamma_3$ between the atoms are highlighted with arrows.
(b) Bilayer graphene electronic bands and allowed vertical inter-band electron-hole transitions.
(c) and (d) Sketch of the expected optical conductivity $\omega\sigma_1$ at the charge neutrality point  $E_F$=0 (c) and at $2E_F=1500 cm^{-1}$ (d).
In both cases a phenomenological broadening of $\sim$ 0.05$\gamma_1$ is used to account for disorder and/or inhomogeneous carrier distribution.
Here we do not show the \textit{1--4} interband excitations, that have onset energy $E_{On}^\textit{1--4}=2\gamma_{1}\sim$ 6000 $cm^{-1}$ and are sensitive to the Pauli blocking for $E_F>\gamma_{1}$.}
\end{figure}

The unit cell of bilayer graphene contains four nonequivalent atoms $A1$, $B1$, $A2$ and $B2$, where letters $A$ and $B$ denote two sublattices in the same layer, while $1$ and $2$ stand for the bottom and top layer [see Fig.~\ref{Fig1}(a)]. 
The Fermi level in graphene lies in the vicinity of the corners of the hexagonal Brillouin zone (also called valleys) known as $K_{+}$ and $K_{-}$.
Due to interlayer coupling, the valence and conduction bands of bilayer graphene split in two subbands [see Fig.~\ref{Fig1}(b)].

The conventional tight-binding Hamiltonian is based on $\pi$-orbitals of carbon atoms (one per atom, four in the unit cell).
We take into account only the in-plane coupling $\gamma_0$, the interlayer coupling $\gamma_1$ between dimers atoms $B1$ and $A2$, and the interlayer coupling $\gamma_3$ between non-dimer orbitals $A1$ and $B2$.
The main term in the Hamiltonian, linear in momentum $\mathbf{p}$ around the valleys reads 
\begin{equation}
\hat{H}_{0} = 
\begin{pmatrix}
\xi v_{3}(\sigma_{x}p_{x}-\sigma_{y}p_{y}) & \xi v \boldsymbol{\sigma}\cdot\mathbf{p} \\
\xi v \boldsymbol{\sigma}\cdot\mathbf{p} & \gamma_{1}\sigma_{x}
\end{pmatrix},
\label{eq:H0}
\end{equation}
where $v=\sqrt{3}a\gamma_0/2\hbar$ is the band velocity, $v_3=\sqrt{3}a\gamma_3/2\hbar$, $\boldsymbol{\sigma}=(\sigma_{x}$,~$\sigma_{y})$ and $\sigma_{x}$, $\sigma_{y}$, $\sigma_{z}$ are the Pauli matrices, $\xi=\pm$ is the valley index.
The basis is constructed using components corresponding to atomic sites $A1,B2,A2,B1$ in the valley $K_{+}$ and $B2,A1,B1,A2$ in $K_{-}$. 
The linear Hamiltonian results in the low energy electronic bands having the parabolic behavior, which transforms into linear at high frequency [see Fig.~\ref{Fig1}(a),(b)]. 
The next order, quadratic in the electron momentum, of the tight-binding Hamiltonian is 
\small
\begin{equation*}
\delta{\hat{H}} \!=\! 
%\mu 
%\begin{pmatrix}
%\frac{v_{3}}{v}[\sigma_{x}(p_{x}^{2} \!-\! p_{y}^{2}) \!+\! 2\sigma_{y}p_{x}p_{y}] & \sigma_{x}(p_{x}^{2}\!-\!p_{y}^{2}) \!-\! 2\sigma_{y}p_{x}p_{y} \\
%\sigma_{x}(p_{x}^{2} \!-\! p_{y}^{2}) \!-\! 2\sigma_{y}p_{x}p_{y} & 0 
%\end{pmatrix},
%
%
%\mu \sigma_{x}(p_{x}^{2}-p_{y}^{2})
%\begin{pmatrix}
%\frac{v_{3}}{v} & 1 \\
%1 & 0 
%\end{pmatrix}
%+
%2\mu \sigma_{y}p_{x}p_{y}
%\begin{pmatrix}
%\frac{v_{3}}{v} & -1 \\
%-1 & 0 
%\end{pmatrix},
%
\mu
\begin{pmatrix}
\frac{v_{3}}{v}(\sigma_{x}Q_{x}-\sigma_{y}Q_{y}) &  \boldsymbol{\sigma}\cdot\mathbf{Q} \\
\boldsymbol{\sigma}\cdot\mathbf{Q} & 0
\end{pmatrix}
\label{eq:dH}
\end{equation*}
\normalsize
where $\mu_{2}=-\frac{v^{2}}{6\gamma_{0}}$, $Q_{x}=p_{x}^{2} - p_{y}^{2}$, $Q_{y}=- 2p_{x}p_{y}$, and $\mathbf{Q}=(Q_{x},Q_{y})$.
Note, that, unlike in monolayer case, the triagonal warping is created by $H_{0}$, and not by $\delta H$.

Because of the vanishing momentum transfer in the Raman process with visible photons, the ERS spectrum will be dominated by vertical interband transitions. These vertical transitions are shown in Fig.~\ref{Fig1}(b) in the band structure of bilayer graphene. 
For illustrative purposes, we first describe the excitation spectrum ignoring Raman matrix element selection rules (which is crucial as we demonstrate later) and take into account the energy and momentum conservation rules only.
In such case, the ERS spectrum does not depend on photon polarizations and is simply given by the imaginary part of the dynamical electronic polarizability $\Pi$, which, in turn, is closely related to the optical conductivity $\sigma_1$ as $\Pi''\propto\omega\sigma_1$~\cite{Shastry1990, Grushin2009}.
The Figs.~\ref{Fig1}(c) and (d) show the frequency dependence $\omega\sigma_1$ for bilayer graphene due to the interband transitions at the charge neutrality point (c) and at finite bias [i.e.\ $E_F\neq 0$] (d) using the theoretical expression of $\sigma_1$, which proved itself to be a good description of the infrared absorption spectrum of bilayer graphene~\cite{AbergelFalko2007,Zhang2008,Kuzmenko2009,Li2009}. 
\par
The \textit{2--3} interband transitions have the lowest energy and their spectrum has an onset $2E_F$, what makes them similar to the one observed in monolayer graphene. For $\gamma_1>E_{F}>0$ the \textit{2--4} interband transitions show an onset at $\gamma_{1}$, while \textit{1--3} transitions are activated from $\gamma_{1}$+2$E_F$ on, as the Pauli blocking is lifted off. 
On the other hand, the \textit{3--4} transitions do not show an onset, but a peak centered approximately at $\gamma_1$, where at finite doping the intensity increases strongly due to increased phase space~\cite{Zhang2008,Wang2008,Li2009,Kuzmenko2009}.

\subsection{Raman processes in bilayer graphene and connection to isospin}

In order to define the Raman selection rules, we need to describe the interaction of the electrons with electromagnetic field.
To achieve this, the canonical momentum should be introduced, $\mathbf{p}-e(\mathbf{A}_I+\mathbf{A}_S)$, where $\mathbf{A}_I$ and $\mathbf{A}_S$ are vector potentials of the incoming and outgoing light, respectively.
We expand the resulting Hamiltonian up to the second order in the vector potential and write down the interaction part:
\begin{equation}
\begin{split}
{\hat{H}_{\textrm{int}}} &= \mathbf{j}\cdot\bigl( \mathbf{A}_{I}+\mathbf{A}_{S}\bigr) + M_{w},
\\
\hat{M}_w &= \frac{e^{2}}{2}\sum_{i,j}A^i_{I}A^j_{S}\partial_{p_{i}}\partial_{p_{j}}\delta\hat{H} ,
\end{split}
\label{eq:AA}
\end{equation}
where $\mathbf{j}= (j_{x},j_{y})$, $j_{i}=-e\partial_{p_{i}}\hat{H}_{0}$ is the current operator and $\hat{M}_w$ is an amplitude of the one-step Raman process.
The two-step process is realized through the subsequent absorption/emission processes described by the first term in Eq.~\eqref{eq:AA}.
Due to the small parameter $v/c$ the excitation energy in the intermediate state is of order of its energy in the finite state and is much smaller then the incoming light frequency $\Omega_I$, making the intermediate state of the whole system virtual.
The virtual absorption process may precede the emission process creating a large excess of energy in the virtual state ($\approx\Omega_I$), or follow the emission process leading to a large deficit of energy in the virtual state ($\approx-\Omega_I$). Thus, they correspond to zero and two-photon intermediate states respectively. 

\begin{figure}
\includegraphics[width=.8\columnwidth]{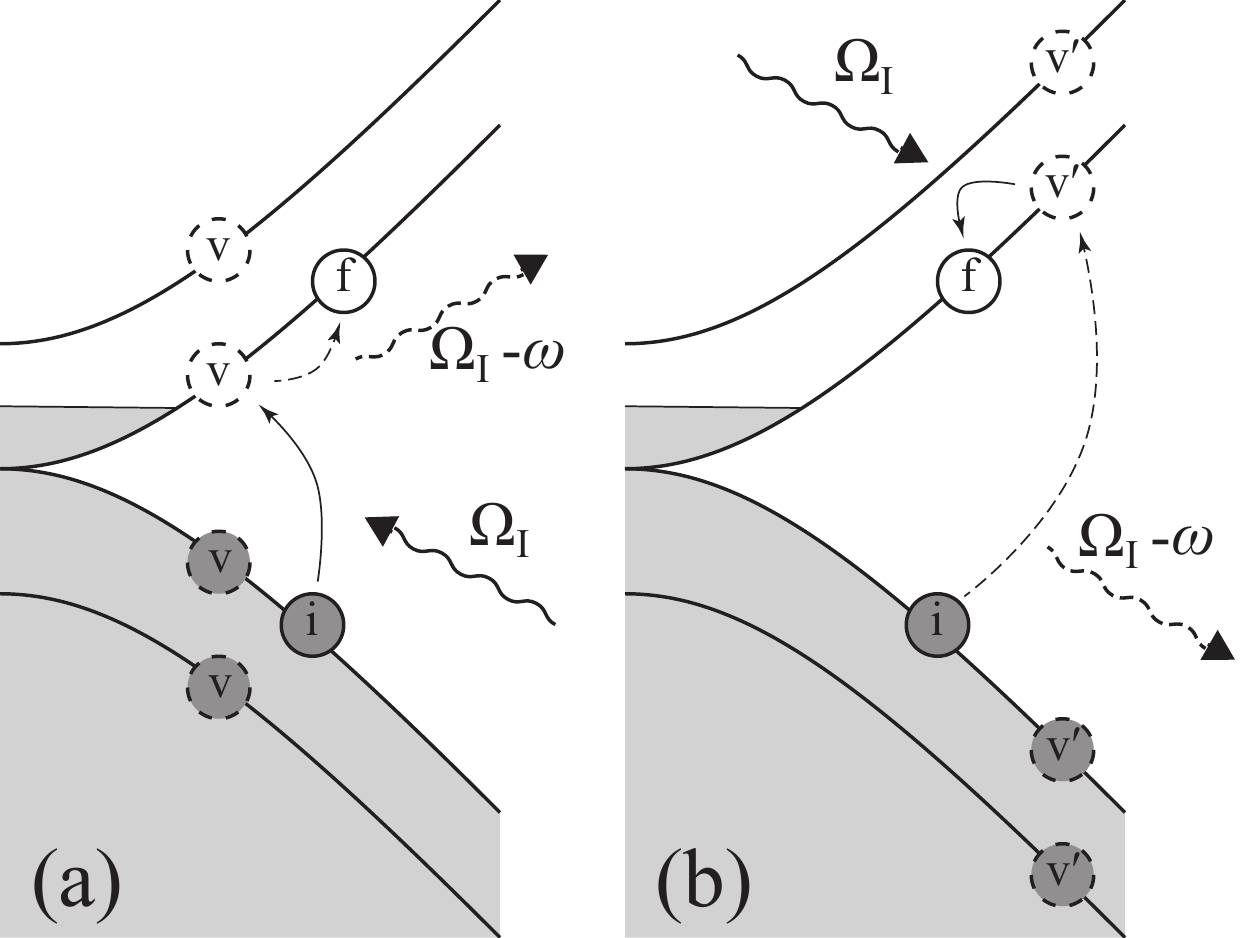}
\caption{\label{fig:transitions} Two-step Raman processes in bilayer graphene.
Indices i, v and f correspond to initial, intermediate (virtual) and final electronic states respectively.
$\Omega_I$ is the incoming light frequency and $\omega$ is the Raman shift.
The absorption (emission) processes are represented by solid (dashed) lines.
They correspond to processes involving zero (a) and two-photon (b) virtual intermediate states.}
\end{figure}

The two kinds of two-step Raman scattering processes are illustrated in Fig.~\ref{fig:transitions}(a)~and~(b).
In each figure the absorption (solid lines) and emission (dashed lines) events can take place in any order.
Note, however, that by swapping absorption and emission processes (within one figure) we also change the sequence of the creation and annihilation operators in the initial (i) and finite (f) electronic states.
Therefore, each of figures (a) and (b) describes two sequences (with direct and reverse order of absorption and emission events) that constitute the Raman scattering event.
The swapping of the photons creation and annihilation operators does not change the expression due to their bosonic nature, while the same swapping for the electron ladder operators results in the minus sign due to their fermionic statistics.
We now calculate explicitly the Raman probability of two-step electronic transitions in bilayer graphene.
For the two possible sequences in the Fig.~\ref{fig:transitions}(a) we get
%
%\begin{widetext}
\begin{multline*}
\sum_{v}\left[
\frac{a_{f}^+a_{v}\,\, a_{v}^{+}a_{i}}{\Omega_I - \epsilon_{v} + \epsilon_{i}} +
\frac{a_{v}^{+}a_{i}\,\,a_{f}^+a_{v}}{-(\Omega_I - \omega) - \epsilon_{f} + \epsilon_{v}}  \right] =\\=
\sum_{v}\left[
\frac{1-f_{v}}{\Omega_I - \epsilon_{v}+\epsilon_{i}} a_{f}^+a_{i}+
\frac{f_{v}}{-\Omega_I + \epsilon_{v}-\epsilon_{i}} a_{i}a_{f}^+ \right] =\\=
\sum_{v}\frac{1}{\Omega_I - \epsilon_{v}+\epsilon_{i}} a_{f}^+a_{i} ,
\end{multline*}
%\end{widetext}
%
where index $v$ denotes all possible momentum-preserving virtual states, and $a$, $a^+$ are fermion destruction and creation operator, respectively. Here we also used the relation $\epsilon_{f}=\epsilon_{i}+\omega$ where $\omega>0$ is the Raman shift, i.e.\ the energy of the electronic excitation created in the system.
Similarly, for the Fig.~\ref{fig:transitions}(b) we get
%\begin{widetext}
\begin{multline*}
\sum_{v'}\left[
\frac{a_{v'}^{+}a_{i}\,\, a_{f}^+a_{v'}}{\Omega_I - \epsilon_{f}+\epsilon_{v'}} +
\frac{a_{f}^+a_{v'}\,\,a_{v'}^{+}a_{i}}{-(\Omega_I-\omega) -\epsilon_{v'} + \epsilon_{i}}  \right] =\\ 
 =\sum_{v'}\left[
\frac{f_{v'}}{\Omega_I - \epsilon_{f}+\epsilon_{v'}} a_{i}a_{f}^+ +
\frac{1-f_{v'}}{-\Omega_I + \epsilon_{f}-\epsilon_{v'}} a_{f}^+a_{i} \right] =\\=
\sum_{v'}\frac{1}{-\Omega_I - \epsilon_{v'}+\epsilon_{f}} a_{f}^+a_{i}.
\end{multline*}
%\end{widetext}
%
By means of the expressions for the current operators, the expression for the Raman amplitude of the two-step process can be simply written as
\begin{multline}
\hat{M}_{D} = 
\bigl( \mathbf{j}\cdot \mathbf{A}_{S}\bigr)[\Omega_{I}-H_{0}+\epsilon_{i}]^{-1}\bigl(\mathbf{j}\cdot\mathbf{A}_{I}\bigr) + \\+
\bigl(\mathbf{j}\cdot\mathbf{A}_{S}\bigr)[-\Omega_{I}+\omega-H_{0}+\epsilon_{i}]^{-1}\bigl( \mathbf{j}\cdot \mathbf{A}_I\bigr).
\label{eq:RDdef}
\end{multline}

The momentum conservation allows the estimation of the components of the Hamiltonian $H_{0}$ as $\omega$ or $\gamma_{1}$, depending whether the intermediate state is in the low or high energy band.
For typical frequency of the visible light ($\sim$ 2 eV) we have $\gamma_1 \ll \Omega_I$, so for small Raman shift, $\omega \ll \Omega_I$,  the denominators can be approximated by $\sim \Omega_I$ and $\sim - \Omega_I$ for the both kinds of Raman processes. 
%In our analysis we neglect the momentum of the photon, since the Fermi wavelength for $E_{F}=800\text{cm}^{-1}$ is $\sim 50\text{nm}$, what is much less than the laser wavelength $\sim 500 \text{nm}$.
%Strong destructive interference effects between the two amplitudes are thus expected to occur if incoming and outgoing photon polarizations are linear and identical.
%However, as we show below, this is not the case if they are orthogonal.

Expanding over the $1/\Omega_I$, the Raman amplitudes for the one-step and two-step processes can be estimated as
\begin{align*}
\hat{M}_{w} &\sim \frac{e^{2}}{\Omega_I} \mu \sim\frac{e^{2}v^{2}}{6 \Omega_I\gamma_{0}},
\\
\hat{M}_{D} &\approx  \frac{e^{2}v^{2}}{\Omega_I^{2}} \!\left\{[j_{x},j_{y}]_{-} (\mathbf{e}_I\times{\mathbf{e}_S})_{z} \!+\! O[\omega/\Omega_I] \!+\! O[\gamma_{1}/\Omega_I] \right\}.
\end{align*}
Here, $\mathbf{e}_I$ and $\mathbf{e}_S$ are the polarizations of the incoming and scattered photons.
Since $\gamma_0 \gg \Omega_I$ the one-step term is small and the ERS signal will be dominated by two-step processes. 
Neglecting also the terms proportional to $v_{3}^{2}$ in the currents commutator ($v_{3}/v \sim 0.1$) we get
\begin{multline}
\hat{M} \approx \frac{e^{2}v^{2}}{\Omega^{2}}\Biggl\{ 
\begin{pmatrix}
\sigma^{z} & 0 \\
0 & \sigma^{z}
\end{pmatrix}
(\mathbf{e}_I\times\mathbf{e}_S)_{z} +\\+\!  O[v_{3}^{2}/v^{2}] \!+\!  O[\omega/\Omega_I] \!+\! O[\gamma_{1}/\Omega_I] \!+\! O[\Omega_I/\gamma_0]
\Biggr\}.
\label{eq:R}
\end{multline}

As we can see from the Eq.~\eqref{eq:R}, the resulting Raman amplitude is dominated by processes in which incoming and outgoing photon polarizations are orthogonal. In terms of the symmetry representation of the point group of graphene, the scattering processes correspond to the $A_2$ symmetry representation~\cite{kashuba2012}. As discussed above, this Raman selectivity can be traced back to the destructive interference effects between the two kinds of two-step processes shown in Fig.~\ref{fig:transitions}.

Interestingly, the dominant $A_2$ symmetric Raman amplitude has a simple connection with the isospin, which in 4-bands model is defined in the basis of the Hamiltonian~\eqref{eq:H0} as
\begin{equation}
\mathbf{S}=\sigma^{0}\otimes \boldsymbol{\sigma}\equiv
\begin{pmatrix}
\boldsymbol{\sigma} & 0 \\
0 & \boldsymbol{\sigma}
\end{pmatrix},
\qquad
\boldsymbol{\sigma} = (\sigma^{x},\sigma^{y},\sigma^{z}),
\label{eq:S}
\end{equation}
where $\sigma^{0}$ is a unity matrix.
This definition is equivalent to the spin definition in the 2-bands model given in Refs.~\onlinecite{Mccann2011,Novoselov2006}.
As one can see, the $A_2$ Raman amplitude is proportional to the $z$ component of the isospin and can be written as $\hat{M}=\frac{e^{2}v^{2}}{\Omega_I^{2}} (\mathbf{e}_I\times\mathbf{e}_S)_{z} S^{z}$.

\subsection{Symmetry of excitations and relation to chirality}
We have shown that the dominant Raman processes have $A_2$ symmetry and transforms like the $z$ component of the isospin, and now investigate to which interband transitions they correspond to.
For this we analyze the structure of the general solution of the Hamiltonian (\ref{eq:H0}), and demonstrate that the dominant Raman processes are excitations between bands symmetric with respect to the $E=0$ line, i.e.\ from band \textit{3} to \textit{2} and from \textit{4} to \textit{1} only. We further show that they correspond to a flip of the isospin orientation, and are thus chiral.

Let us consider the structure of the eigenstates of the Hamiltonian~\eqref{eq:H0} in the new basis $A1,A2,B2,B1$.
The Hamiltonian $H_{0}$ then takes form
\begin{equation*}
H_{0} = 
\begin{pmatrix}
 0 & h \\
h^{*} & 0 
\end{pmatrix} ,
\qquad
h=
\begin{pmatrix}
v_{3}\mathbf{p}^{*} & v \mathbf{p} \\
v \mathbf{p} & \gamma_{1} 
\end{pmatrix} 
\end{equation*}
where $\mathbf{p}=p_{x}+ip_{y}$.
The solutions of the Sch\"odinger equation $H_{0}|\psi\rangle=E|\psi\rangle$ are represented by eigenvectors
\begin{equation}
\begin{split}
|\pm,a/b\rangle &= 
\frac{1}{\sqrt2}
\begin{pmatrix}
|a/b\rangle \\
\pm\lambda_{a/b}^{-1/2}h^{*}|a/b\rangle 
\end{pmatrix},
\\
hh^{*}|a/b\rangle &= \lambda_{a/b}|a/b\rangle,
\end{split}
\end{equation}
with eigenvalues $E=\pm \sqrt{\lambda_{a/b}}$ such that $|a/b\rangle$ and $\lambda_{a/b}$ are eigenvectors and eigenvalues of the matrix $hh^{*}$, correspondingly.
Note that $\lambda_{a/b}$ are real and positive.
These eigenstates coincide with the bilayer bands: $|-,a\rangle\equiv|1\rangle$, $|-,b\rangle\equiv|2\rangle$, $|+,b\rangle\equiv|3\rangle$, and $|+,a\rangle\equiv|4\rangle$.
In this new basis the isospin and Raman amplitude [in the limit of Eq.~\eqref{eq:R}] are equal to
\begin{equation}
\mathbf{S}= \boldsymbol{\sigma}\otimes\sigma^{0},
\qquad
\hat{M}\propto S^{z}\equiv
\begin{pmatrix}
\sigma^{0} & 0 \\
0 & -\sigma^{0} 
\end{pmatrix}.
\end{equation}
Since all eigenvectors are orthogonal, the Raman amplitude $\hat{M}\propto S^{z}$ couples the states with the opposite sign of the energy
\begin{equation}
S^{z}|\pm,a/b\rangle = |\mp,a/b\rangle,
\end{equation}
which also means that the expectation value of the isospin vector $\langle n |\mathbf{S}| n \rangle$ lies in the graphene's plane, since the $z$-component of its the expectation value is zero.
In terms of the band transitions shown in Fig.~1(b), the dominant Raman scattering process invokes the transitions \textit{1--4} and \textit{2--3} only. All other transitions have vanishingly small Raman amplitude as described above. 
The anti-commutation relation $M S^{x/y}=-S^{x/y} M$ indicates that the initial and final electronic states involved in the Raman process (see Fig.~\ref{fig:transitions}) have the same momentum, but possess opposite isospins. They have thus opposite signs of the spin projection onto momentum, i.e. opposite chirality in graphene's language.

\section{Electronic Raman scattering in bilayer graphene: experiments}

In this section we test experimentally the theoretical predictions made above in a bilayer graphene device.
With the broader aim of establishing Raman scattering as a selective probe of chiral excitations in graphene systems, we have two specific objectives: to show that the dominant Raman processes have $A_2$ symmetry, and test whether they can be ascribed to \textit{2--3} interband transitions at low energy. 
\subsection{Methods}
\begin{figure}  
  \includegraphics [width= \columnwidth]{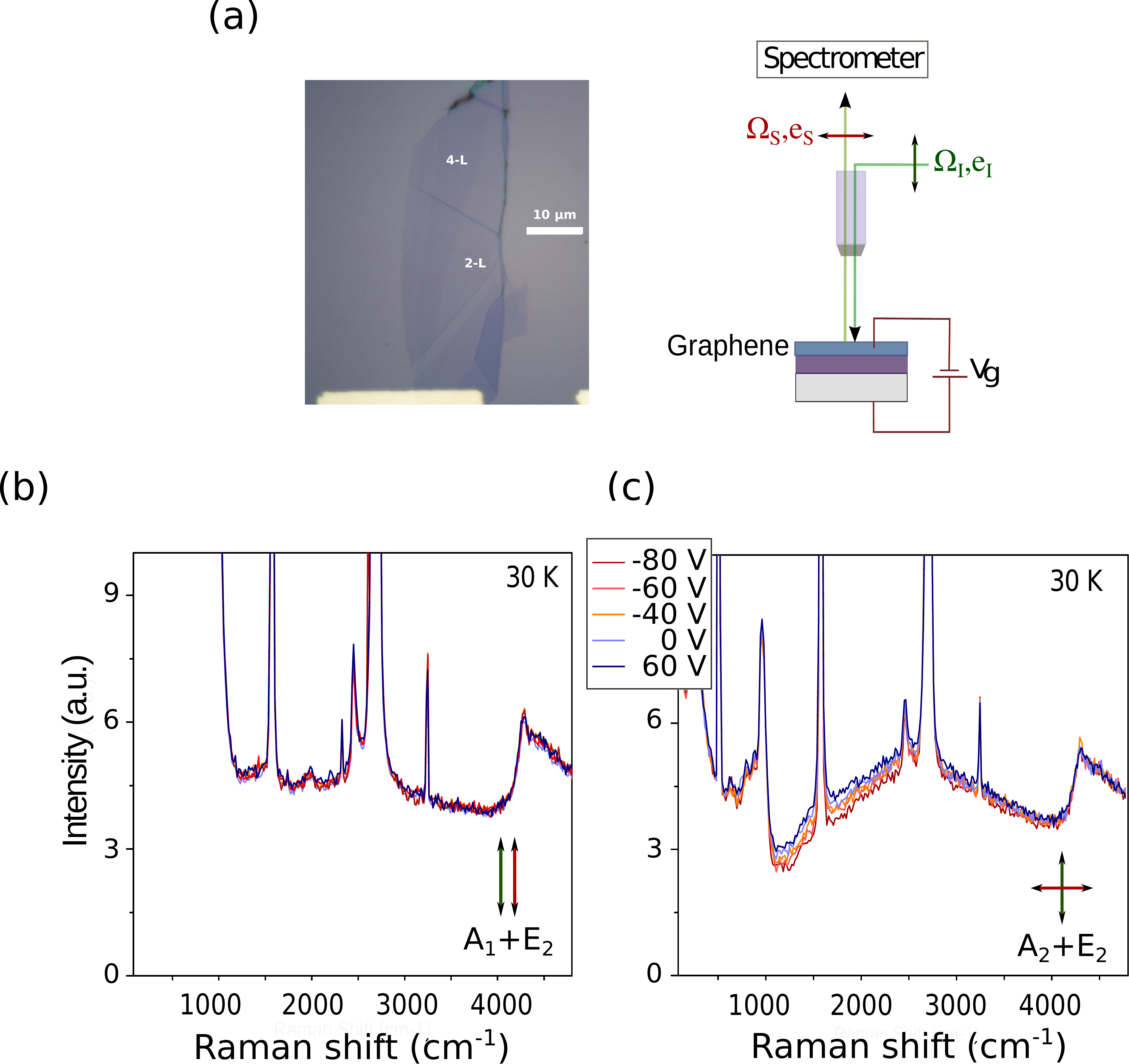}
   \caption
   { \label{Fig3} 
   (a) Optical microscope image of bilayer graphene device, and schematic drawing of the set-up. $\mathbf{e}_{I,S}$ and $\Omega_{I,S}$ are the polarizations and frequencies of incoming and scattered photons. The Raman shift is defined as $\omega=\Omega_I-\Omega_S$. (b) and (c) Polarization resolved Raman continuum of bilayer graphene recorded at $T=$30$K$ at five different gate voltage values for parallel (a) and cross (b) polarizations of incoming and scattered photons.
   }
\end{figure}

The studied graphene samples were produced by exfoliation of natural graphite and characterized by phononic Raman spectroscopy. The study of 2D band ($\sim$ 2600 cm$^{-1}$) and M band (1700-1800 cm$^{-1}$) features allow to identify the sample thickness unambiguously, following the Ref.~\onlinecite{Nguyen2014}.
The sample shown in Fig.~\ref{Fig3}(a) consists of regions with bilayer and quadrilayer Bernal-stacked graphene.
Electrical contacts were first fabricated using e-beam lithography and Pd deposition on an oxidized Si wafer with a SiO$_2$ thickness of $\sim 500 \text{nm}$).  

The graphene flake was then positioned on top of the Si/SiO$_2$ device with contacts using a dry transfer technique under an optical microscope. In such a structure the doped Si substrate acts as a back gate [see Fig.~\ref{Fig3}(a)].
We use a back gate to distinguish the ERS contribution from the other sources of background signal in the measured spectra.
The graphene device was first characterized by studying the gate voltage evolution of the G band phonon energy and linewidth. The charge neutrality point ($E_F=0$) was found at $50\pm 10 \text{V}$. The polarization resolved Raman scattering measurements were performed using a home-built micro-Raman setup in a backscattering configuration.
The $\lambda=532\text{nm}$ ($2.33\text{eV}$) excitation line of a Diode Pumped Solid State (DPSS) Laser  was  focused  onto  the  sample using a long-working-distance 100X objective lens with N.A.=0.8.
The laser spot size was $ \leq 1 \text{$\mu$m}$ and  all  measurements  were  performed with an incident laser power less than $1\text{mW}$ and in the vacuum chamber ($P\leq 10^{-5}\text{mbar}$) of a low temperature optical cryostat. The lowest cold finger temperature achieved was $\sim 30\text{K}$.
The excitation beam and the collected signal were linearly polarized in order to identify the symmetry of the Raman active excitations. 

\subsection{Gate dependent electronic Raman spectrum of bilayer graphene}

Figures \ref{Fig3}(b) and (c) show Raman continuum of bilayer graphene recorded at $T=30\text{K}$ as a function of the applied back gate voltage, in parallel (b) and cross (c) polarizations which probe all excitations belonging to $A_{1}+E_{2}$ and $A_{2}+E_{2}$ symmetry representations of bilayer graphene, respectively~\cite{Riccardi2016}.
The spectra exhibit sharp peaks corresponding to the well-known Raman-active optical phonons of graphene layers.
Here we focus on the underlying broad continuum, which partly originates from the ERS processes described above.
In parallel polarization configuration, the Raman spectrum does not show any sign of gate dependence, indicating that it likely arises from non-ERS background, or resonant higher-order ERS processes~\cite{Hasdeo2014}, as it was found for a monolayer device~\cite{Riccardi2016}.
On the other hand, the cross polarizations spectra exhibit a clear gate effect, as expected for non-resonant ERS signal coming from low energy interband excitations, with a gradual and partial suppression of intensity as the gate voltage increases~\cite{Mucha2010,Hasdeo2014}.
The fact that the gate dependent signal is only seen in cross-polarization indicates a dominant contribution to the ERS continuum from electronic excitations having $A_{2}$ symmetry representation as expected theoretically above.
Note that the gate dependent signal strength is at least $20\%$ of the measured signal in cross-polarization configuration and has $A_{2}$ symmetry.
As we will show below, the partial suppression of the $A_2$ ERS continuum upon increasing gate voltage is well-reproduced by considering the Pauli blocking effects on the chiral \textit{2--3} interband transitions.
   
In order to isolate the gate dependent ERS signal, we have subtracted the phonon peaks and normalized  the resulting spectra with the one recorded at the estimated  charge neutrality voltage ($E_F=0$). The gate dependent signal can then be discussed in terms of the ratio $R(\omega, V_{G})$ :
\begin{equation}
R(\omega, V_{G})=\frac{I(\omega, V_{G})}{I(\omega, V_{CN})}=\frac{I_{0}+I_{ERS}(\omega, V_{G})}{I_{0}+I_{ERS}(\omega, V_{CN})}
\end{equation}
Here $I_{ERS}(\omega, V_{G})$ is the gate-dependent ERS bilayer continuum intensity and $I_0$ the gate independent background signal, which we assumed to be weakly frequency dependent. In the Fig.~\ref{Fig4}(a) we show $R$ for several gate voltages.
When the Fermi energy is moved away from charge neutrality point, the ratio $R$ decreases and the onset energy of this suppression gradually shifts at higher frequencies, as expected for Pauli blocking effects on interband transitions (see inset of Fig.~\ref{Fig4}(a)).

\subsection{Theoretical modeling of the spectra}

\begin{figure} 
  \includegraphics [width=.75 \columnwidth]{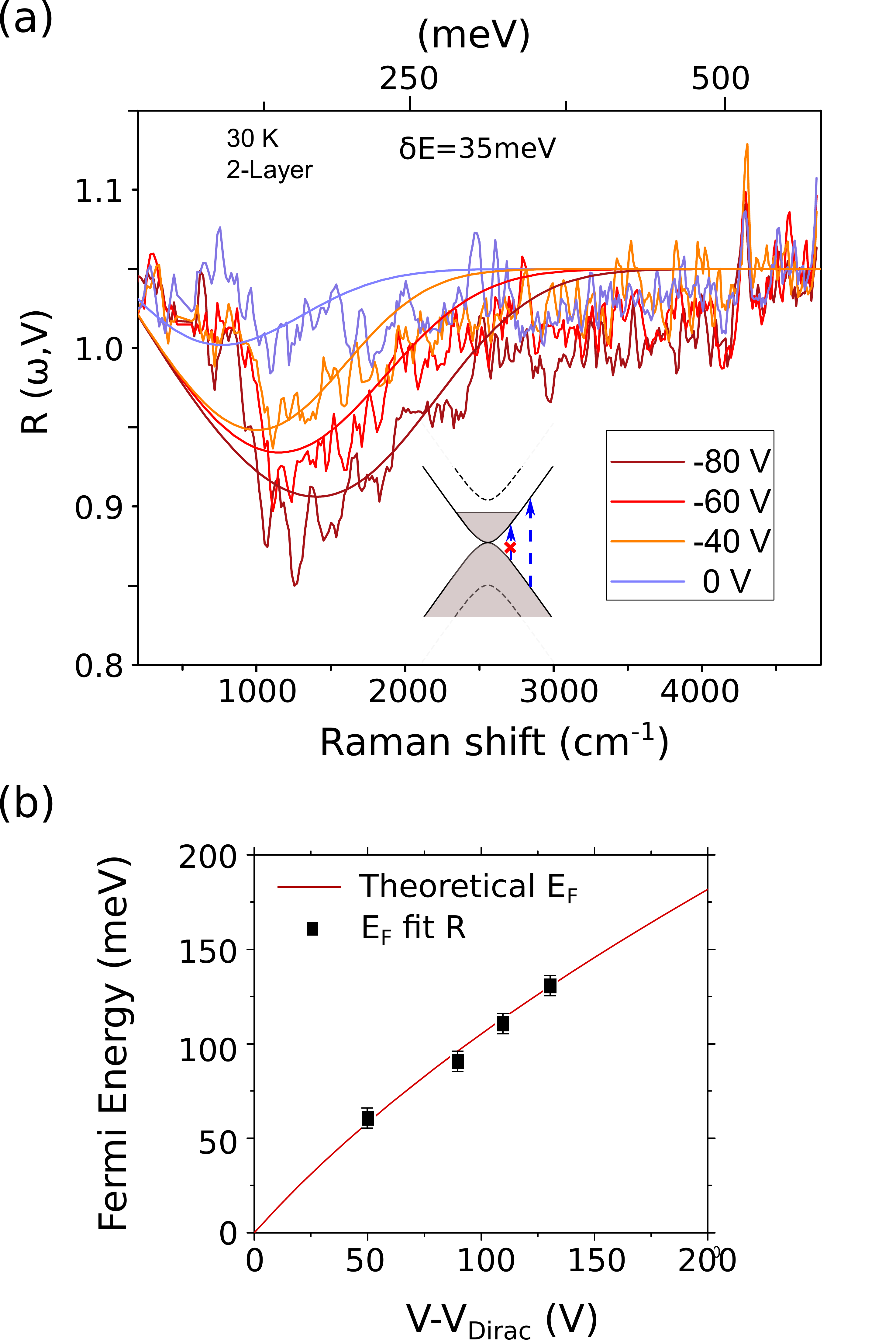}
   \caption
   {\label{Fig4} (a) Experimental and theoretical $R(\omega, V_{G})$ at 30 $K$ for the cross polarizations spectra. The theoretical plots are obtained by only considering vertical 2-3 interband transitions, with a Gaussian distribution of Fermi energy with $\delta E_{F}\sim 35 meV$ .
   (b) Adjusted values of $E_F$ in function of the gate voltage, superposed to the theoretical curve \ref{nef}.}
\end{figure}

To analyze the gate dependent ERS signal we compute the expected theoretical $R$ profiles. We only take into account the Raman signal coming from the \textit{2--3} transitions since, as shown theoretically above, they are expected to dominate the ERS spectrum at low Raman shift, $\omega < 2\gamma_1$. In this case, the ERS intensity can be analytically computed by using the expression of the Raman amplitude $\hat{M}$ and the Fermi golden rule. Extending the results of Ref.~ \cite{Mucha2010} to finite temperature and using the corresponding dispersion relation, we obtain the following frequency dependence:
\begin{equation}
I_{ERS}^{2-3}\propto \frac{\omega+\gamma_1}{\pi (\hbar v_F)^2} \left[ f\left( -\frac{\hbar \omega}{2}\right)  - f\left( \frac{\hbar \omega}{2}\right) \right] 
\end{equation}
where $f(\epsilon)=[1+e^{\frac{\epsilon -E_F}{k_B T}}]^{-1}$ is the Fermi-Dirac distribution.
Note that in the low energy or parabolic limit, $\omega \ll \gamma_1$, we recover the constant ERS continuum found in Ref.~\onlinecite{Mucha2010}. The adjustable parameters in this simple model are $E_F$, $\delta E_F$, a distribution of Fermi energy due to the inhomogeneous charge doping.
Here $I_{0}$ is an additional constant describing an energy independent background.
The Fig.~\ref{Fig4}(a) demonstrates a good agreement of experimental and theoretical results for ratio $R$.
In particular, as shown in Fig.~\ref{Fig4}(b), the $E_F$ values used for the fits (black dots) are very close to the ones expected by using the relationship between the Fermi energy $E_F$ and the charge density $N$ deduced from the gate voltage and the estimated device capacitance $C\sim 60\text{aF/$\mu$m}^2$:
\begin{equation}
E_F=\frac{-\gamma_1+\sqrt{\gamma_1^2+4\pi N(\hbar v_F)^2}}{2}
\label{nef}
\end{equation}

The variation of Fermi energy assumed for the theoretical curves is $\delta E = 35\text{meV}$, is consistent with previous estimations for a similar bilayer graphene device \cite{Yan2008}. Moreover, it is consistent with the energy distribution coming from the fits performed on the gate evolution of the G band in the same device (see below). 

\subsection{Link with G band renormalization}

\begin{figure}
\includegraphics [width=.75 \columnwidth]{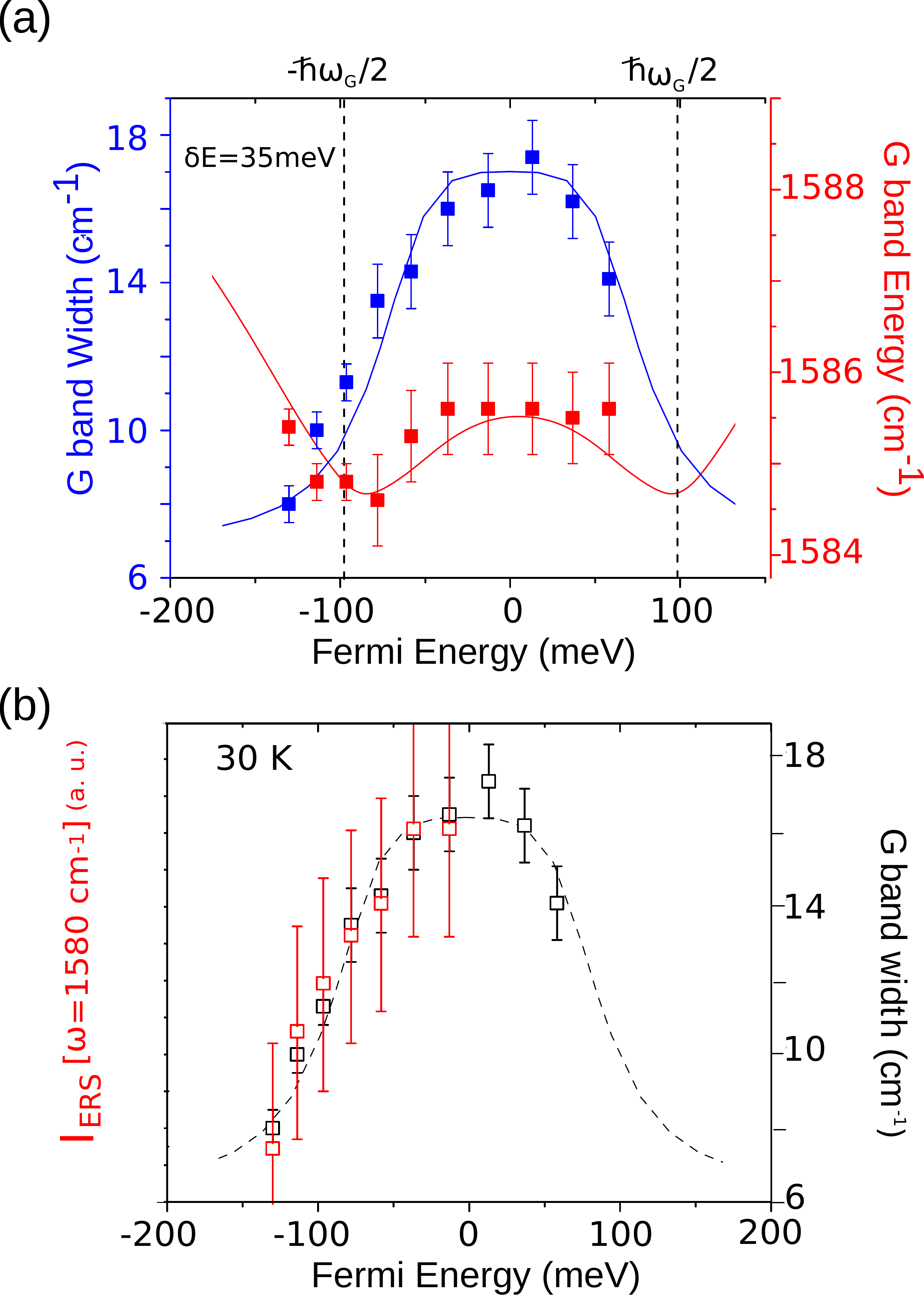}
\caption{\label{Fig5}
(a) Evolution of G band linewidth (blue dots) and energy (red dots) of the bilayer graphene sample as a function of Fermi energy.
The Fermi energy was deduced from the gate voltage using the same device capacitance as in the case of the ERS spectra.
Theoretical plots are obtained with the same variation of the Fermi energy, $\delta E_{F}\sim 35\text{meV}$, as the $R(\omega, V_{G})$ theoretical fits [see Fig.~\ref{Fig4}(c)]. 
One of the two anomalies is clearly identified at $E_F = -\hbar \omega_G /2$.
(b) Evolution of the ERS intensity at 1580 $cm^{-1}$ (red dots) and of the G band linewidth (black dots) as a function of the Fermi energy, and theoretical expectation (dashed line).}
\end{figure}

The fitting parameters used for the ERS spectrum can be independently cross-checked by looking at the G band renormalization under gate voltage. The coupling between phonons and low-energy electrons-hole pairs in graphene devices has been studied both theoretically~\cite{Ando2006,Ando2007,Neto2007} and experimentally~\cite{Pisana2007,Yan2007,Yan2008}.
The Figure~\ref{Fig5}(a) shows the G band energy and width as a function of the Fermi energy in our device.
When the Fermi energy crosses half of the phonon frequency ($E_F=\pm\hbar\omega_G/2$), the G band displays an anomalous softening and its linewidth drops sharply~\cite{Yan2008}.
In experimental measurements, the profile of this anomaly strongly depends on the inhomogeneity of the charge distribution providing an independent estimate of its value. As shown in fig. \ref{Fig5}(a) the value of the variation of Fermi energy $\delta E_F$ assumed in the fit of $R$ is perfectly consistent with the one coming from the G band energy and linewidth renormalizations.
\par
In addition, we also note that the linewidth renormalization of the G band is directly connected to the vertical electron-hole pairs excitations with energy $\omega_G$~\cite{Ando2006, Ando2007}.
At this energy only \textit{2--3} transitions are allowed and they have mainly chiral $A_{2}$ character, as already discussed for the ERS spectra. As such they should not couple to the G band phonon, which has $E_{2}$ symmetry. However, taking into account trigonal warping, the \textit{2--3} vertical interband transitions acquire a small but finite $E_{2}$ component, which allows their coupling to the G band, as discussed by e.g.\ Basko~\cite{Basko2008}.
The superposition of the ERS intensity taken at $\omega \sim 1580\text{cm}^{-1}$, and the G band linewidth at several values of the Fermi energy, is shown in Figure \ref{Fig5}(b).
The similar evolutions of the two quantities indicate that they are indeed both connected to \textit{2--3} interband transitions.

\subsection{Contrast between optical conductivity and ERS in bilayer graphene}
\begin{figure} 
\includegraphics [width= .75 \columnwidth]{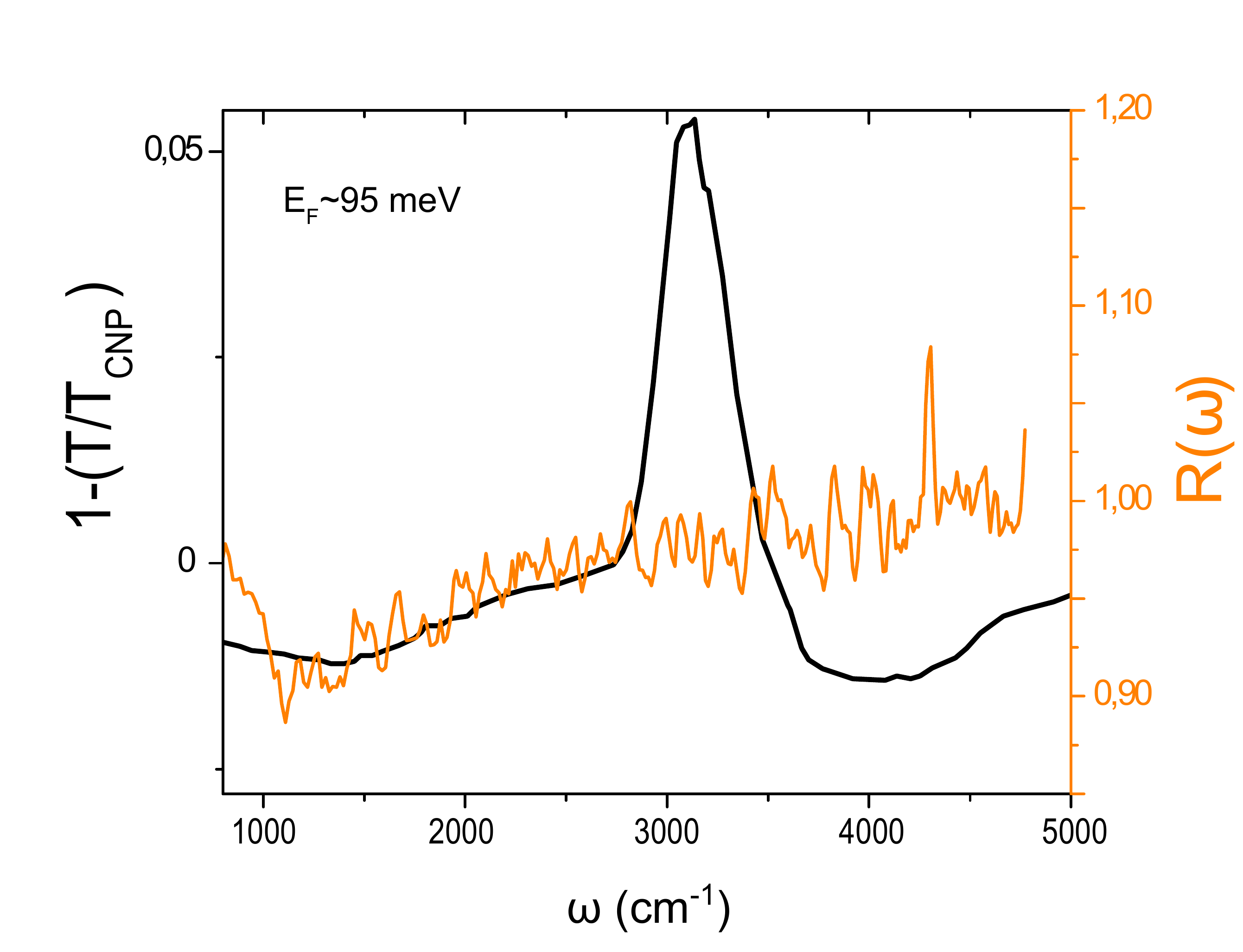}
\caption{\label{Fig6}
Comparison of $R(\omega, V_{G})$ with normalized infrared transmission measurements $(1-T/T_{CN})$ where $T_{CN}$ is the transmission for a Fermi energy at the charge neutrality point.
Both were performed on bilayer graphene devices~\cite{Li2009} with approximately the same Fermi level shift.
The $1-(T(V)/T_{CN})$ spectrum is dominated by a peak at around $3000\text{cm}^{-1}$, with a magnitude that increases together with the gate voltage.
By contrast, the ERS intensity at this frequency is essentially independent of gate voltage.}
\end{figure}

Unlike in the monolayer case, the difference between optical conductivity and Raman response of bilayer graphene is substantial. In monolayer graphene a single set of vertical interband transition is possible and selection rules play a marginal role.
In such case ERS and optical conductivity provide essentially the same information and are related via a simple relation, $I_{ERS}=\omega\sigma_1$.
As shown above, the ERS spectrum of bilayer graphene is well reproduced by taking into account the $A_{2}$ chiral transitions between the bands \textit{2} and \textit{3} only, neglecting all other interband transitions. By contrast, all interband transitions contribute to optical conductivity.
The Fig.~\ref{Fig6} illustrates the striking difference of ERS and infrared transmission measurements for the case of bilayer graphene. It shows a superposition of $R(\omega, V_{G})$ at $-40\text{V}$ and a normalized infrared transmission data taken from Ref.~\onlinecite{Li2009}, which closely mirrors the gate-induced changes in the optical conductivity of bilayer graphene.
Both set of data were taken on a bilayer graphene device with a comparable Fermi energies, $E_F \sim 95\text{meV}$.
The infrared spectrum is dominated by a peak at $\sim 3000\text{cm}^{-1}$ arising from non-chiral \textit{3--4} transitions at $\gamma_1$ as expected from the theoretical prediction of the optical conductivity. This prominent peak is masking the broader contribution from \textit{2--3} chiral transitions in the infrared spectrum. By contrast, the peak at $\gamma_1$ is absent in the ERS spectrum, confirming that non-chiral excitations are filtered out in the ERS process.

\section*{Conclusion}

In this paper we have presented an electronic Raman scattering study of a bilayer graphene device at varying gate voltage.
The spectra show a remarkable selectivity of the Raman probe on the interband excitations that require the inversion of the electron chirality. Theoretically, this selectivity is ascribed to the fact that the dominant electronic Raman processes of bilayer graphene belong to the $A_2$ symmetry, which includes the chiral electronic transitions only.
We demonstrated that the processes in other symmetries are suppressed due to the strong destructive interferences effects in the Raman amplitudes.
This selectivity contrasts with infrared transmission measurements and opens interesting venues for the use of Raman scattering as a selective probe of isospin and chiral degrees of freedom in graphene, topological materials, and other 2D crystals.

\appendix*

\section{\label{apx:chirality} Chirality, helicity, and spin-momentum locking}

In our work we use the terminology accepted by the graphene community referring to the spin-momentum locking phenomenon as chirality~\cite{Mccann2011,Mccann2006,Mccann2006a,Novoselov2006}.
Nevertheless, it is important to introduce clearer definitions, in order to clarify which properties we probe in the experiment. The chiral excitations, by definition, are the solutions of the Hamiltonian, for which the parity transformation (a flip in the sign of \emph{one} spatial coordinate, which is equivalent to a reflection through a line in 2D, and through a plane in 3D) cannot be compensated by a rotation~\cite{Peskin1995}. The valleys in monolayer graphene play a crucial role in making the excitations chiral, since the reflexion exchanges the K and K$^\prime$ points.
To compare with the high-energy physics models, the solutions of the 2D Dirac Hamiltonian $\boldsymbol{\sigma}\cdot\mathbf{p}$ constructed by means of $2\times2$ Pauli matrices $\sigma^{x/y}$ are non-chiral.
Meanwhile, the same Hamiltonian in 3D (with extra term $\sigma^{z}p_{z}$), contains only chiral solutions, since it itself is not symmetric with respect to the parity (which in 3D is equivalent to inversion---the flipping in the sign of all three spatial coordinates). The chirality of monolayer graphene excitations is preserved also at higher energies, where the trigonal warping is relevant and the rotations degrade from SO(2) to C$_{6}$. These statements about chirality are also correct for bilayer graphene.

If one follows the particle physics terminology, the spin-momentum locking property of monolayer graphene referred to in graphene community as chirality is actually called \emph{helicity}~\cite{Peskin1995}, which is defined as the projection of the angular momentum or spin onto the direction of momentum. Indeed, the helicity of the state in the monolayer graphene takes the discreet values of $+1$ and $-1$ denoting also the band which this state belongs to.
In case of the bilayer graphene this simple definition fails. For the particular case of low-energy excitations in bilayer graphene, the Ref.~\onlinecite{Novoselov2006} defines helicity as $\boldsymbol{\sigma}\cdot(p_{x}^{2}-p_{y}^{2}, 2p_{x}p_{y})/p^{2}$. In case of the four-bands model of bilayer graphene this definition can be generalized, defining helicity as a \emph{continuous} map of the momentum direction onto the iso-spin direction.

As we stated above, in the Raman scattering process the momentum of the excitation is kept fixed, but its isospin direction is changed to the opposite one. The excitation changes the band crossing to the state which is mirror-symmetric with the respect to the $E=0$ line.  Thus, the electron and hole states of the Raman-induced pair have the opposite projections of the isospin onto momentum. This meets the definition of the term ``Raman chiral excitations'' if we use the terminology established by the graphene community.

% If you have acknowledgments, this puts in the proper section head.
%\begin{acknowledgments}
% put your acknowledgments here.
%\end{acknowledgments}

%\section*{References}
\bibliographystyle{apsrev4-1}
%\bibliography{Bilayer_Chiral}

%merlin.mbs apsrev4-1.bst 2010-07-25 4.21a (PWD, AO, DPC) hacked
%Control: key (0)
%Control: author (72) initials jnrlst
%Control: editor formatted (1) identically to author
%Control: production of article title (-1) disabled
%Control: page (0) single
%Control: year (1) truncated
%Control: production of eprint (0) enabled
%

\end{document}